\documentclass[conference]{IEEEtran}
\usepackage{amsmath,amssymb,bm,booktabs,balance,graphicx,xcolor}
\usepackage[hidelinks]{hyperref}
\newcommand{\R}{\mathbb R}
\newcommand{\Hh}{\mathcal H_t}
\newcommand{\Sc}{\mathcal S_t}

\begin{document}
\title{Who Gets the Grid Capacity? Allocating Dynamic Access Among Multiple Virtual Power Plants}
\author{
    \IEEEauthorblockN{Nakul Sreekanth, Jun He, Andrew L. Liu, and S. Sivaranjani}
    \IEEEauthorblockA{Intelligent Electric Power Grid Lab, Purdue University\\
    West Lafayette, IN, USA\\
    \{nsreekan, he184, andrewliu,  sseetha\}@purdue.edu}
}
\maketitle
\begin{abstract}
Virtual power plants (VPPs) trade in the same wholesale market and share distribution lines and transformers, and uncoordinated, profit-maximizing VPPs can push a shared transformer past its thermal limit or drive a feeder voltage out of band. The utility, which only observes the front-of-the-meter net injection at each bus, must keep the network safe from that net injection alone. We allocate a 24-hour active and reactive operating envelope to each VPP by a Benders decomposition: the utility sets a trial envelope, each VPP solves its private convex portfolio problem, and the returned value and dual gradient become supporting cuts that rebuild its value function while its schedule, state, and cost stay private. The envelope carries the transformer's thermal state across hours, bounds a joint conditional value-at-risk over forecast uncertainty, and is re-optimized each hour with the current hour committed. On a reduced IEEE-13 feeder with real load and prices, an active-power-only implementation lowers the transformer's peak-temperature CVaR from 254 to 97\,$^\circ$C at a 28\% value give-up and holds the transformer below its limit where nameplate access breaches it.
\end{abstract}

\section{Introduction}
A VPP aggregates distributed energy resources such as solar generation, batteries, and electric vehicles (EVs) \cite{ruan2024}. With vehicle-to-grid (V2G) technology, EVs can both draw power from and supply power to the grid \cite{kempton2005}. When independent VPPs respond to the same wholesale price, they may charge or export at the same time. Trades that are feasible for each portfolio can therefore overload shared equipment or violate voltage limits. The utility must keep the network within its limits while each VPP chooses its own trades. This coordination problem arises in practice. Australia's Project EDGE brought Mondo, Discover Energy, and Rheem/Combined Energy Technologies into AusNet's distribution area \cite{edge}, and SA Power Networks and Tesla demonstrated dynamic export limits during wholesale and frequency-control trading \cite{sapn}.

The utility must also account for how these trades affect the network over time. Transformer temperature depends on both current and prior loading \cite{thermal,c5791}. Solar generation initially reduces import current, but large exports can drive current back up, and simultaneous battery or EV charging can sustain heavy loading across several hours. The same trade can therefore have different thermal consequences depending on when it occurs. This motivates us to allocate network access over a rolling 24-hour horizon and carry the transformer's thermal state forward across hours and days.

\subsection{Related Work}
We review prior studies that design feasible and fair operating envelopes, allocate network access among aggregators, and use decomposition to coordinate schedules across the grid.

\emph{Network-feasible and fair envelopes.} \cite{sapn} reports a single-aggregator mechanism trial in Australia, and \cite{edge} studies economic envelopes for one aggregator. \cite{wickramasinghe} surveys the state of the art and identifies fairness and coordination across aggregators as open problems, while \cite{ruan2024} reviews data-driven VPP management more broadly. \cite{petrou} minimizes the squared distance between each limit and intended operation, and \cite{alam} allocates envelopes in two stages with technical and equitable objectives. \cite{liu} builds a robust hyperrectangle with proportional fairness. \cite{koirala} uses a chance-constrained stochastic OPF, and \cite{yi} builds a Gaussian chance-constrained fair envelope. \cite{mahmoodi} splits capacity by a one-shot Kornai--Liptak decomposition, \cite{gholami} uses per-VPP boundaries from a small LP, and \cite{quessongo} adds storage recourse over multiple periods. These methods treat the prosumer as passive and allocate one period at a time, so the transformer's thermal state, which couples hours, stays outside the allocation. \cite{petrou} notes that its framework is ``not applicable in the presence of multiple aggregators,'' and \cite{liu} identifies multi-aggregator coordination as an open problem.

\emph{Economic access allocation.} \cite{chen} runs a forward auction that clears network access to competing DER aggregators, taking reported bids at face value and leaving bid formation open. \cite{attarha} reports a matching two-aggregator trial with the same value-aware clearing. \cite{tsaousoglou} designs a mechanism that makes truthful cost bids dominant, and \cite{zhang} shows that a VCG payment restores efficient dispatch among strategic learning agents. These papers price a flexibility service. Allocating private envelopes to several competing VPPs remains open.

\emph{Decomposition and intertemporal scheduling.} \cite{li} uses Benders decomposition to operate networked microgrids with the utility while keeping each model private. \cite{yuan} decomposes the TSO-DSO dispatch with an affine generalized bid function, and \cite{li2026} learns envelopes end to end with an input-convex network. These methods coordinate a single agreed schedule or a learned policy.

Across these works, the allocation runs one period at a time for a single aggregator, or treats reported bids as given. In our work, we instead allocate a 24-hour envelope to several competing VPPs through Benders cuts that carry only value and gradient, keep the transformer's thermal state across hours, and evaluate the envelope under forecast error.

\subsection{Contributions}
We develop a method that allocates operating envelopes to multiple VPPs sharing a feeder to maximize their total trading value. Our contributions are three-fold. First, we account for each VPP's intertemporal device constraints, feeder voltage limits, and line flow limit. Each hour, we re-optimize the full 24-hour active- and reactive-power envelopes and commit only the current hour, so that future allocations can adapt to updated forecasts. Second, we derive Benders value and feasibility cuts from each VPP's private convex portfolio problem. The utility coordinates access using reported values and gradients, while each VPP keeps its device model private. Third, we distinguish worst-case operation within the awarded envelopes from uncertainty in the forecasts and use a joint CVaR constraint \cite{cvar} to limit the risk of network violations due to forecast errors.

\section{Rolling-Horizon Coordination Model}
\label{sec:model}
At issue time $t$, we coordinate delivery hours $h\in\Hh=\{t,\ldots,t+23\}$ with duration $\Delta=1$~h. We use a finite scenario tree with leaves $s\in\Sc$, probabilities $\pi_s>0$, and $\sum_s\pi_s=1$ to represent hourly nonparticipating load, photovoltaic (PV) availability, ambient temperature, electric-vehicle (EV) availability, and exogenous prices. We observe first-hour information and impose nonanticipativity: decisions sharing an observed history must agree. We fix the tree and forecasts throughout the cut iterations at $t$.

\subsection{VPP's Problem and Private Cuts}
VPP $i\in\{1,\ldots,N\}$ connects at bus $n(i)$. We define positive $p_{ihs},q_{ihs}$ as active and reactive injection and negative values as withdrawal. Each connection point receives separate bounds. For a single connection, we stack four hourly bounds over 24 hours into $a_i\in\R^{4|\Hh|}=\R^{96}$, with
\begin{equation}
a_{ih}=(\underline p_{ih},\overline p_{ih},\underline q_{ih},\overline q_{ih}), \text{ where } \begin{cases}
 \underline p_{ih}\le p_{ihs}\le\overline p_{ih},\\
 \underline q_{ih}\le q_{ihs}\le\overline q_{ih}.
 \end{cases}
\label{eq:box}
\end{equation}
We apply these bounds to every scenario. A positive lower active-power bound requires a minimum export in that hour, so we accept it only after the VPP certifies feasibility. We include reactive-power bounds because both active and reactive injections affect voltage and current. Rectangular envelopes allow us to protect the network against independent use of the allocated ranges. We retain device-level inverter capability constraints in the private problem. Coupled polygons, apparent-power limits, ramps, and cumulative-energy bounds could be incorporated into a more detailed access product.

We collect the VPP's decisions and states over the scenario tree in $x_i\in\R^{n_i^x}$. This vector includes PV output $g_{ihs}$; battery charging power $c_{ihs}$, discharging power $d_{ihs}$, and stored energy $e_{ihs}$; the EV, heating, water-heating, and appliance variables introduced below; inverter reactive powers; and net injections $p_{ihs}$ and $q_{ihs}$. We use $n_i^x$ for the dimension of this vector. For PV and stationary storage, we impose
\begin{equation}
\begin{split}
&0\le g_{ihs}\le\widehat g_{ihs},\\
&e_{i,h+1,s}=e_{ihs}+\Delta\left(\eta_c c_{ihs}-\frac{d_{ihs}}{\eta_d}\right),\\
&\underline e_i\le e_{ihs}\le\overline e_i,\quad 0\le c_{ihs}\le m_{ihs}\overline c_i,\\
&0\le d_{ihs}\le(1-m_{ihs})\overline d_i,
\end{split}
\label{eq:bat}
\end{equation}
where the input parameters are defined as follows: $\widehat g_{ihs}$ is available PV power; $\eta_c,\eta_d\in(0,1]$ are charging and discharging efficiencies; $\underline e_i,\overline e_i$ are energy limits; and $\overline c_i,\overline d_i$ are charging and discharging power limits. We specify the initial energy $e_{it}$ and terminal reserve $e_{i,t+24,s}\ge e_i^{\rm end}$ as inputs. We also prescribe the binary charging-mode indicator $m_{ihs}\in\{0,1\}$ before allocation, consistently with nonanticipativity. Fixing this indicator keeps the private problem convex and supports the value cuts derived below.

For EV $j$, we denote plug-in availability by $\alpha_{jhs}$, charging power by $v_{jhs}$, V2G discharge power by $w_{jhs}$, battery energy by $e^v_{jhs}$, and energy withdrawn for driving by $d^v_{jhs}$. We impose
\begin{equation}
\begin{split}
&0\le v_{jhs}\le\alpha_{jhs}\overline v_j,\quad 0\le w_{jhs}\le\alpha_{jhs}\overline v_j,\\
&e^v_{j,h+1,s}=e^v_{jhs}+\eta_j v_{jhs}\Delta-\frac{w_{jhs}\Delta}{\eta_j}-d^v_{jhs},\\
&\underline e^v_j\le e^v_{jhs}\le\overline e^v_j,\qquad e^v_{j,h^{\rm dep}_{js},s}\ge e^{\rm req}_{js}.
\end{split}
\label{eq:ev}
\end{equation}
We specify power limit $\overline v_j$, efficiency $\eta_j$, and energy limits $\underline e^v_j,\overline e^v_j$. Availability, driving withdrawals, arrival energies, departures $h^{\rm dep}_{js}$, and required energies $e^{\rm req}_{js}$ are scenario inputs. EVs are bidirectional: V2G discharge $w_{jhs}$ shares the charging power limit and efficiency, and is available only while plugged in. For heat-pump power $b_{jhs}$ and indoor temperature $T_{jhs}$, we use
\begin{equation}
\begin{split}
&T_{j,h+1,s}=\rho_jT_{jhs}+(1-\rho_j)T^{\rm amb}_{hs} +\beta_j b_{jhs}+\gamma_{jhs},\\
&\underline T_{jh}\le T_{jhs}\le\overline T_{jh},\qquad 0\le b_{jhs}\le\overline b_j .
\end{split}
\label{eq:heat}
\end{equation}
Here $\rho_j\in(0,1)$ is thermal retention, $\beta_j$ is the signed heating/cooling coefficient, and $\gamma_{jhs}$ represents other heat gains. We specify comfort bounds $\underline T_{jh},\overline T_{jh}$ and power limit $\overline b_j$. For water-heater power $b^z_{jhs}$ and stored heat $z_{jhs}$, we use
\begin{equation}
\begin{split}
&z_{j,h+1,s}=\rho^z_jz_{jhs}+\eta^z_j b^z_{jhs}\Delta-d^z_{jhs},\\
&\underline z_j\le z_{jhs}\le\overline z_j,\qquad 0\le b^z_{jhs}\le\overline b^z_j ,
\end{split}
\label{eq:water}
\end{equation}
Here $\rho^z_j\in(0,1]$ is heat retention, $\eta^z_j>0$ is heat delivered per unit electricity, and $d^z_{jhs}$ is hot-water heat withdrawal. We specify storage limits $\underline z_j,\overline z_j$, power limit $\overline b^z_j$, initial states $T_{jt},z_{jt}$, and terminal comfort/reserve bounds.

For a divisible appliance or pool-pump job, we constrain power $f_{jhs}$ by $0\le f_{jhs}\le\overline f_j$ within the unfinished eligible hours $\mathcal W_j$ and set it to zero elsewhere, where $\overline f_j$ and $\mathcal W_j$ are inputs. For an in-horizon deadline with input energy requirement $E_j^{\rm job}$, we require $\sum_{h\in\mathcal W_j}f_{jhs}\Delta=E_j^{\rm job}$. For a later deadline, given initial remaining requirement $E^{\rm rem}_{jt}$, we track the terminal remaining energy $E^{\rm rem}_{j,t+24,s}$ through
\begin{equation}
\begin{split}
&0\le E^{\rm rem}_{j,t+24,s} =E^{\rm rem}_{jt}-\sum_{h\in\Hh\cap\mathcal W_j}f_{jhs} \Delta \le\overline f_j \Delta|\mathcal W_j\setminus\Hh|.
\end{split}
\end{equation}
This bound preserves completion capacity conditional on future network access. Noninterruptible jobs require integer variables. With customer base active load $L^P_{ihs}$, we write
\begin{equation}
\begin{split}
p_{ihs}=&g_{ihs}+d_{ihs}-c_{ihs}-L^P_{ihs}\\
&-\sum_j(v_{jhs}-w_{jhs}+b_{jhs}+b^z_{jhs}+f_{jhs}).
\end{split}
\label{eq:balance}
\end{equation}
We sum each device term over the corresponding devices in VPP $i$, suppressing the owner index on $j$. Reactive injection is $q_{ihs}=\sum_jq^{\rm inv}_{jhs}-L^Q_{ihs}$, with known or affine reactive demand $L^Q_{ihs}$ and inverter limits $(p^{\rm inv}_{jhs})^2+(q^{\rm inv}_{jhs})^2\le(\overline S_j)^2$. We equate inverter active flow $p^{\rm inv}_{jhs}$ to its connected devices' combined net output, with apparent-power limit $\overline{S}_j$. Unit-wise, powers use MW/Mvar and electrical energies MWh.

Let $\mathcal X_{i,t}$ denote the constraint sets of device, balance, initial/terminal, job, and nonanticipativity constraints, applying state bounds through $t+24$. With exogenous nodal price $\lambda_{ihs}$ (\$/MWh), convex interval cost $C_{ihs}$ (\$), and convex terminal cost $\Psi_i$ (\$), we define
\begin{equation}
\begin{split}
&V_{i,t}(a_i) = \\
& \max_{x_i\in\mathcal X_{i,t}} \sum_s\pi_s(\sum_{h\in\Hh} (\lambda_{ihs}p_{ihs}\Delta-C_{ihs}(x_i))-\Psi_i(x_{i,t+24,s}))\\
& \text{s.t. } A_i x_i\le b_i+B_i a_i.
\end{split}
\label{eq:vpp}
\end{equation}
Here $x_{i,t+24,s}$ is the terminal-state subvector of $x_i$ in scenario $s$. The constraint stacks \eqref{eq:box}. With $r_i = 4|\Hh||\Sc|=96|\Sc|$ scenario-stacked envelope rows, we have $A_i\in\R^{r_i\times n_i^x}$, $B_i\in\R^{r_i\times96}$, and $b_i\in\R^{r_i}$.

We assume wholesale-market eligibility and price-taking behavior. We settle net charging at the modeled price and count each energy sale once. In the baseline, we use $C_{ihs}=\kappa_i(c_{ihs}+d_{ihs})\Delta$, where $\kappa_i\ge0$ is the battery-throughput cost, and set $\Psi_i=0$ while retaining the stated terminal reserves. Our settlement covers energy transactions; service payments, bidding, and imbalance settlement would add terms.

For a feasible trial $a_i^k$, we obtain optimal multipliers $\mu_i^k\ge0$ for the stacked envelope inequalities. We use the maximization Lagrangian $F_i(x_i)-\mu_i^\top(A_ix_i-b_i-B_ia_i)$, where $F_i$ is the objective in \eqref{eq:vpp}. Under strong duality with an attained dual optimum, we construct the upper supporting cut for the concave value function:
\begin{equation}
\begin{split}
&w_i\le V_{i,t}(a_i^k)+(B_i^\top \mu_i^k)^\top (a_i-a_i^k).
\end{split}
\label{eq:optcut}
\end{equation}
The master estimate is $w_i\in\R$. We use supergradient $B_i^\top\mu_i^k$ directly because the objective already includes scenario probabilities.

For an infeasible envelope, we solve phase I:
\begin{equation}
\begin{split}
&\varphi_i(a_i)=\min_{x_i\in\mathcal X_{i,t},\,\omega_i\in\R_+^{r_i}} \{\mathbf1^\top\omega_i:A_ix_i\le b_i+B_ia_i+\omega_i\}.
\end{split}
\label{eq:phase}
\end{equation}
We scale rows to common per-unit tolerances and use consistently scaled matrices and multipliers. For optimal $\nu_i^k\ge0$ and $\varphi_i(a_i^k)>0$, we add
\begin{equation}
\begin{split}
&\varphi_i(a_i^k)-(B_i^\top \nu_i^k)^\top (a_i-a_i^k)\le0.
\end{split}
\label{eq:feascut}
\end{equation}
This supporting lower bound on convex $\varphi_i$ excludes the failed proposal. We require $\mathcal X_{i,t}\ne\varnothing$ and valid optimal dual solutions. Cuts keep device records local but can disclose marginal economics; formal privacy and truthful reporting remain unestablished.

For a charging-only fleet needing an additional $E$~MWh by its deadline, with efficiency $\eta$, feasibility requires
\begin{equation}
\begin{split}
&-\Delta\sum_h\underline p_{ih}\ge E/\eta .
\end{split}
\end{equation}
The sum covers charging hours before the deadline. Phase I derives such coupled restrictions for the full portfolio.

\subsection{Utility Problem, Risk Constraints, and Hourly Updates}
We maximize joint VPP value using a balanced, radial, lossless linear power-flow model with buses $\mathcal N$, branches $\mathcal E$, fixed topology, and substation squared voltage $v_0$. We convert network powers to per-unit values and transform envelope coordinates and cut slopes consistently to preserve dollar-valued objectives.

We set $D_{\ell n}=1$ when bus $n$ is downstream of branch $\ell$, and $M_{ni}=1$ when $n=n(i)$, with zero entries otherwise. Vectors $L^P_{hs},L^Q_{hs}$ contain nonparticipating bus loads. Reactive controls $u_{hs}\in\R^{|\mathcal N|}$ satisfy $\underline u\le u_{hs}\le\overline u$, componentwise ramps $|u_{hs}-u_{h-1,s}|\le\overline r$, measured initial output, and nonanticipativity. We set unavailable equipment's bounds to zero and fix discrete tap/capacitor settings. Stacking the VPP bounds at hour $h$, we obtain
\begin{equation}
\begin{split}
&P^-_{hs}=D(L^P_{hs}-M\overline p_h),\\
&P^+_{hs}=D(L^P_{hs}-M\underline p_h),\\
&Q^-_{hs}=D(L^Q_{hs}-u_{hs}-M\overline q_h),\\
&Q^+_{hs}=D(L^Q_{hs}-u_{hs}-M\underline q_h).
\end{split}
\label{eq:flow}
\end{equation}
We set $R_{n\ell},X_{n\ell}$ to branch $\ell$'s resistance and reactance when it lies on the path to bus $n$, and zero otherwise. For nonnegative line parameters,
\begin{equation}
\begin{split}
&v^-_{hs}=v_0\mathbf1-2(RP^+_{hs}+XQ^+_{hs}),\\
&v^+_{hs}=v_0\mathbf1-2(RP^-_{hs}+XQ^-_{hs}).
\end{split}
\label{eq:voltage}
\end{equation}
We introduce nonnegative bounds $\widehat P_{\ell hs}\ge\pm P^-_{\ell hs},\pm P^+_{\ell hs}$ and $\widehat Q_{\ell hs}\ge\pm Q^-_{\ell hs},\pm Q^+_{\ell hs}$, and impose
\begin{equation}
\begin{split}
&W_{\ell hs}\ge\widehat P_{\ell hs}^2+\widehat Q_{\ell hs}^2.
\end{split}
\label{eq:W}
\end{equation}
With minimum squared voltage $\underline v$ and current limit $\overline I_\ell$, $W_{\ell hs}\le\underline v\overline I_\ell^2$ conservatively enforces ampacity when voltage is within bounds. These bounds cover every simultaneous injection combination in the envelopes, with conservatism from independent flow bounds.

For a transformer on branch $\ell_0$ with rated current $I_R$, we set $y_{hs}\ge W_{\ell_0hs}/(\underline v I_R^2)$ to bound squared per-unit loading when voltage is within bounds. We model top-oil temperature $O_{hs}$, winding rise $G_{hs}$, and hotspot bound $H_{hs}$ as
\begin{equation}
\begin{split}
&O_{h+1,s}=\rho_o O_{hs}+(1-\rho_o)(T^{\rm amb}_{hs}+a_o+b_o y_{hs}),\\
&G_{h+1,s}=\rho_w G_{hs}+(1-\rho_w)(a_w+b_w y_{hs}),\\
&H_{hs}\ge O_{h+\delta,s}+G_{h+\delta',s},\quad \delta,\delta'\in\{0,1\}.
\end{split}
\label{eq:thermal}
\end{equation}
We apply the last inequality to all four endpoint combinations. The retention factors are $\rho_o=e^{-\Delta/\tau_o}$ and $\rho_w=e^{-\Delta/\tau_w}$, where $\tau_o,\tau_w>0$ are time constants. We calibrate nonnegative coefficients $a_o,b_o,a_w,b_w$ to upper-bound the steady-state heating curves over a declared loading range $0\le y\le\overline y$.

One construction uses the concave curves $F_o(y)=\Delta O_R[(1+\ell_Ry)/(1+\ell_R)]^{n_o}$ and $F_w(y)=\Delta G_Ry^{n_w}$, with $0<n_o,n_w\le1$. At a fixed tangent point $y_*>0$, we set $b_j=F'_j(y_*)$ and $a_j=F_j(y_*)-b_jy_*$ for $j\in\{o,w\}$. Here $\ell_R$ is the rated load-loss to no-load-loss ratio, and $\Delta O_R$ and $\Delta G_R$ are rated temperature rises. Concavity ensures that the tangents upper-bound the curves. We obtain these parameters from transformer data and update the measured or estimated initial thermal states each hour.

Each first-order thermal state remains between its endpoints under constant hourly inputs. The endpoint combinations in \eqref{eq:thermal} therefore bound the within-hour hotspot when the ambient and loading inputs also bound subhourly peaks. We repeat the model for each monitored transformer.

For each leaf scenario, we introduce $Z_s\in\R$ to bound the largest normalized signed violation:
\begin{equation}
\begin{split}
&Z_s\ge(\underline v-v^-_{nhs})/\sigma_v,\quad Z_s\ge(v^+_{nhs}-\overline v)/\sigma_v,\\
&Z_s\ge(W_{\ell hs}-\underline v\overline I_\ell^2)/\sigma_{I\ell},\\
&Z_s\ge(H_{hs}-\overline H)/\sigma_H\qquad\forall n,\ell,h .
\end{split}
\label{eq:Z}
\end{equation}
Here $\overline v,\overline H$ are squared-voltage and hotspot limits, and all $\sigma$ parameters are fixed positive engineering scales. For tail probability $\epsilon\in(0,1)$, we impose joint conditional value-at-risk (CVaR) \cite{cvar}:
\begin{equation}
\begin{split}
&\zeta+\epsilon^{-1}\sum_s\pi_s\xi_s\le0,\; \xi_s\ge Z_s-\zeta,\; \xi_s\ge0,\;\zeta\in\R.
\end{split}
\label{eq:cvar}
\end{equation}
For example, $\epsilon=0.05$ limits the average signed violation in the worst $5\%$ probability mass to zero. We retain these signed values because taking their positive part would require zero violation in every positive-probability leaf. The CVaR constraint implies a joint chance bound under the specified discrete distribution.

In the observed first hour, we additionally require every violation expression in \eqref{eq:Z} to be nonpositive. Our thermal constraints address peak hotspot temperature. To impose a no-additional-aging requirement, we would also need a cumulative aging budget relative to a specified no-VPP reference case.

We define the registered-access set as
\begin{equation}
\begin{split}
\mathcal A^{\rm reg}=\{&a:\ p_i^{\min}\le\underline p_{ih}\le\overline p_{ih}\le p_i^{\max},\\
&q_i^{\min}\le\underline q_{ih}\le\overline q_{ih}\le q_i^{\max}\quad\forall i,h\}.
\end{split}
\label{eq:reg}
\end{equation}
The outer bounds are connection-agreement inputs. At iteration $k$, we solve the master problem below with all cuts collected at time $t$:
\begin{equation}
\begin{split}
&\max\quad \sum_i w_i\\
&\text{s.t.}\quad a\in\mathcal A^{\rm reg}, 
w_i\le\overline V_{i,t}, \forall i, \text{and } \eqref{eq:optcut}, \eqref{eq:feascut}, \eqref{eq:flow}-\eqref{eq:cvar}.
\end{split}
\label{eq:master}
\end{equation}
with decision variables $(a,w,u,\widehat P,\widehat Q,W,y,O,G,H,Z,\zeta,\xi)$ and $P^\pm,Q^\pm,v^\pm$ as affine expressions. Constraints \eqref{eq:flow}--\eqref{eq:cvar} include magnitude and loading bounds, first-hour hard limits, reactive-control bounds and ramp limits, and nonanticipativity, and they fix the initial thermal states at their measured values. We obtain $\overline V_{i,t}$ from \eqref{eq:vpp} without envelope rows. We require nonempty private domains, bounded values, strong duality, and dual attainment. The master problem is convex, with quadratic-cone and affine constraints. Minimum widths or fairness criteria require additional policy choices.

At each issue time, we refresh states, forecasts, and value bounds. We solve \eqref{eq:master}, evaluate trial envelopes at all VPPs, and append their cuts. We retain the best fully feasible candidate, with value $\mathrm{LB}$, and use the master optimum as $\mathrm{UB}$. We stop when $\mathrm{UB}-\mathrm{LB}\le\delta_{\rm abs}+\delta_{\rm rel}\max(1,|\mathrm{LB}|)$ for declared tolerances, then award the stored incumbent. Without an incumbent, we issue no new access. Finite termination is unestablished for general conic subproblems; cuts are discarded after state, forecast, or price changes unless re-established.

Each hour we re-optimize the entire 24-hour access trajectory against the current forecast and implement the first-hour actions; the remaining hours are tentative and are re-decided at the next issue time, so the current hour is the firm commitment. Re-optimizing each hour lets the utility track realized conditions, and a safe envelope is re-derived at each hour; a product that fixes multi-hour access would require a certified terminal continuation set and backup policy. If no safe envelope exists at an hour, the utility applies the best available curtailment and records the resulting constraint violation.

\section{Numerical Illustration}
\label{sec:results}

\subsection{Setup}
We apply the model in Section~\ref{sec:model} to a nine-bus reduction of the IEEE 13-bus feeder with a 5-MVA substation transformer. We connect a PV/battery portfolio at bus 680 and a battery/EV portfolio at bus 671. Each stationary battery is rated at 0.25 MW/1 MWh, and base PV capacity is 1 MW. The EV fleet contains 100 vehicles, each with a 75-kWh battery and a 7-kW charger. We use California ResStock loads, clear-sky solar profiles, and CAISO NP15 prices, with loads and prices drawn from different years. We forecast demand using the previous day's values for the same hours and forecast day-ahead LMPs using seasonal ARIMA$(1,0,1)\times(1,0,1)_{24}$. We assume unity power factor and use linear voltage sensitivities and scalar branch limits. This gives conservative voltage and flow limit estimates which the conic $P^\pm,Q^\pm,W$ bounds would tighten.

\begin{figure}[t]
\centering
\includegraphics[width=\columnwidth]{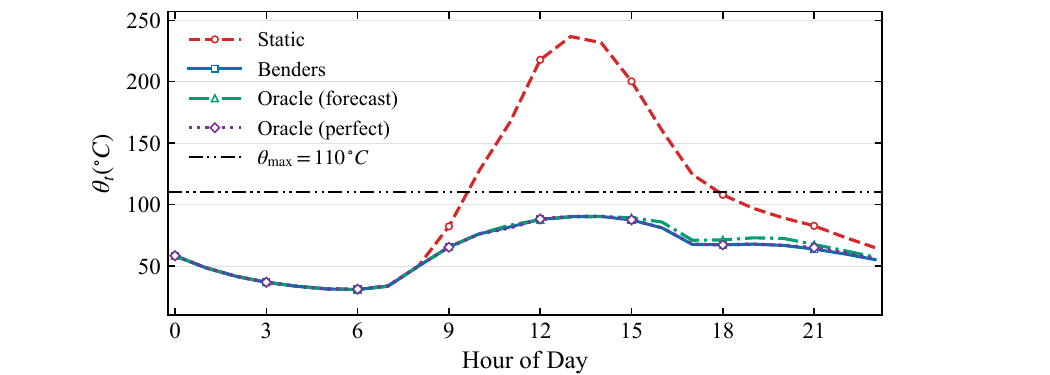}
\caption{Transformer hotspot temperature on one clear day at 12-MW PV, using the single-state model
\eqref{eq:thermal_numerical}. Static nameplate access produces a peak near $237^\circ$C; Benders and
both centralized oracles clip the export to the line rating and hold the peak near $90^\circ$C,
coinciding because the rating caps all of them.}
\label{fig:export}
\end{figure}

The temperatures in Table~\ref{tab:results} and Fig.~\ref{fig:export} follow the single-state
hotspot model:
\begin{equation}
\theta_{h+1}=\rho_\theta\theta_h+(1-\rho_\theta)\left[T^{\rm amb}_h+\Delta\theta(S_h/S_R)^2\right],
\label{eq:thermal_numerical}
\end{equation}
which is a simplification of the two-state model \eqref{eq:thermal} to one state, with $S_R=5$ MVA,
$\Delta\theta=100^\circ$C, and $\tau_\theta=3$ h.

We compute the average daily aggregate trading value (\$/day) over days 150--200 for the export rows and days 150--250 for the load rows. Each row reports the realized value from \eqref{eq:vpp} and the empirical 95\% CVaR \cite{cvar}, from \eqref{eq:cvar}, of the 3 signed violations in \eqref{eq:Z}: the daily peak hotspot, the largest positive voltage-band violation, and the largest branch-rating exceedance $\max_{e,h}(|S_{eh}|-\overline S_e)_+$.

\begin{table}[t]
\caption{Mean daily value and empirical 95\% CVaR statistics. Export rows use the hourly re-solve
over days 150--200; load rows use the day-ahead allocation over days 150--250.}
\label{tab:results}\centering\footnotesize
\setlength{\tabcolsep}{3pt}
\begin{tabular}{@{}llrrrr@{}}\toprule
Case & Allocation & Value & Temp. & Volt. & Overload\\
 & & (\$/day) & ($^\circ$C) & (p.u.) & (MVA)\\\midrule
12-MW PV & Nameplate & 2287.6 & \textbf{253.5} & \textbf{0.038} & \textbf{4.59}\\
 & Coordinated & 1654.3 & 96.5 & \textbf{0.010} & 0\\
1.34$\times$ load & Nameplate & 381.5 & \textbf{113.0} & 0 & 0\\
 & Coordinated & 379.3 & 109.9 & 0 & 0\\
\bottomrule
\end{tabular}
\par\smallskip\raggedright\scriptsize Temperature uses \eqref{eq:thermal_numerical}; voltage and
overload are positive excursions. Limits: $110^\circ$C hotspot, 0.95--1.05 p.u. voltage, and the
study's branch MVA ratings.
\end{table}

Figure~\ref{fig:export} shows one clear day at 12-MW PV: the nameplate envelope lets the PV export its full output and the reverse flow overheats the transformer, while the coordinated envelope clips the export to the line rating and holds the transformer near 90\,$^\circ$C. For the export rows, the utility re-optimizes \eqref{eq:master} each hour on the latest forecast and commits the current hour, and the value and temperature are evaluated on the realized dispatch, where each VPP re-solves \eqref{eq:vpp} at the realized LMP inside its envelope \eqref{eq:box}. Nameplate access drives the peak-temperature CVaR of \eqref{eq:Z} to 253.5\,$^\circ$C through the 5.26-MVA line; the coordinated envelope clips export to the line rating, holding the transformer at 96.5\,$^\circ$C (a 62\% reduction) at a 27.7\% loss in realized value \eqref{eq:vpp}. The flow limit CVaR falls from 4.59 MVA to zero, and the voltage CVaR remains at 0.010 p.u.

The load row solves \eqref{eq:master} once for the day, when the transformer is the binding element. At $1.34\times$ load the nameplate envelope breaches the hotspot limit of \eqref{eq:thermal_numerical} (113.0\,$^\circ$C CVaR), while the coordinated envelope holds the peak at 109.9\,$^\circ$C, just below the limit, for a value cost under 1\%. Beyond this scale the base demand plus EV charging in \eqref{eq:ev} exceeds the transformer rating, and therefore, the coordinated allocation has no feasible point and the utility applies the best available curtailment.

\section{Conclusion}
In our work, we allocate operating envelopes to independent VPPs sharing a feeder. The utility uses Benders value and feasibility cuts to coordinate access while each VPP keeps its device model private. The formulation enforces the transformer hotspot, feeder voltage limits, and line flow limits, as well as uses a joint CVaR constraint to limit the risk of network violations under forecast uncertainty. Each hour, we re-optimize the full 24-hour allocation and commit only the current hour. Future work includes further validation of the allocator, establishing conditions for truthful bidding, and extending the method to firm multi-hour access with a certified continuation policy.

\end{document}